\documentclass[intlimits,twoside,a4paper]{article}
\usepackage{amsmath,amssymb}
\usepackage{graphicx}
\usepackage{wrapfig}
\usepackage[T2A]{fontenc}
\usepackage[cp1251]{inputenc}
\usepackage{cmpj}

\issue{2011}{14}{3}{33001}

\doinumber{10.5488/CMP.14.33001}

\title{Some simple results for the properties of polar fluids}

\author[D. Henderson]{D. Henderson\thanks{E-mail: doug@chem.byu.edu}}
\authorcopyright{D. Henderson, 2011}

\address{Department of Chemistry and Biochemistry, Brigham Young University, Provo UT 84602}

\date{Received March 7, 2011, in final form May 1, 2011}

\begin{document}
\maketitle
\begin{abstract}

The author's lecture notes concerning the correlation functions and the thermodynamics of a simple polar fluid are summarized.  The emphasis is on the
dipolar hard sphere fluid and the mean spherical approximation and on the relation of these results to the Clausius-Mossotti and Onsager formulae for the dielectric constant.  Previous excerpts from these lecture notes, Condens. Matter Phys., 2009, \textbf{12}, 127; ibid., 2010, \textbf{13}, 13002, have contained results that were not widely known.  It is hoped that this third, and likely final, excerpt will prove equally helpful by gathering several results together and making these more widely available and recording a few new results.


\pacs 05.20.-y, 61.20.Gy, 61.20.Ja, 64.30.+t, 77.22.Ch
\end{abstract}

\maketitle

\section{Introduction}

This paper is dedicated to Yura Kalyuzhnyi on the occasion of his sixtieth  birthday; it is the result of the beautiful work of Michael Wertheim on hard spheres and dipolar hard spheres that has inspired the author and many others, including Yura and his colleagues in Lviv.  There is little in this paper, drawn from the author's lecture notes, that is not well-known to Michael but, perhaps, lesser mortals will benefit from this collection of results and those in two previous papers \cite{1,2} taken from the author's lecture notes.

Dipolar hard spheres are a simple representative molecular fluid and polar fluid.  For a canonical ensemble, the well known definition for the h-particle correlation function for a simple atomistic fluid of $N$ molecules in a volume $V$ is easily generalized for a molecular fluid,
\begin{equation}
g(1\cdots  h)=\frac{V^hN!}{N^h(N-h)!}\cdot \frac{1}{Q_N}\int\exp[-\beta U(1\cdots  N)]\rd{\bf{r}}_{h+1}\cdots  \rd{\bf{r}}_{N}\rd\Omega_{h+1}\cdots \rd \Omega_{N},
\end{equation}
where
\begin{equation}
U(1\cdots  N)=\sum_{i<j=1}^{N}u(R_{ij},\Omega_i,\Omega_j)
\end{equation}
is the energy of the system, $\beta=1/kT$ ($T$ is the temperature and $k$ is the Boltzmann constant), $Q_N$ is the configurational partition function given by
\begin{equation}
Q_N=\int \exp[-\beta U(1\cdots  N)]\rd {\bf r}_1\cdots  \rd {\bf r}_N\rd \Omega_1\cdots  \rd \Omega_N\,,
\end{equation}
and $R_{ij}=|{\bf r}_i-{\bf r}_j|$ is the distance between the centers of a pair of molecules, $i$ and $j$, whose centers are located at ${\bf r}_i$ and ${\bf r}_j$.  The ``volume'' elements, $\rd \Omega$, are normalized so that $\int \rd \Omega=1$.  Thus,
\begin{equation}
\rd \Omega=\frac{\sin\theta \rd \theta \rd \phi}{4\pi}\,.
\end{equation}
The function $u(R_{ij},\Omega_i,\Omega_j)$, that can be written as $u(ij)$ for brevity, is the intermolecular potential between a pair of molecules.

The following notation is employed.  A function, such as $g(ij)$, that depends upon the orientation of each member of a pair of particles, $i$ and $j$, is denoted by the presence of each of the indices of the two particles in the argument of the function.   After integration over the orientations of the two particles, a function that depends only on the scalar separation, $R$, of the two molecules results.  This spherically averaged function is denoted by the subscript $0$.  Thus, the radial distribution function (RDF) is given by
\begin{equation}
g_0(R)=\int g(12)\rd \Omega_1\rd \Omega_2=\langle g(12) \rangle.
\end{equation}
The Ornstein-Zernike (OZ) equation becomes
\begin{equation}
h(12)=c(12)+\rho\int h(13)c(23)\rd {\bf r}_3\rd \Omega_3\,,
\end{equation}
where $\rho=N/V$, $h(12)=g(12)-1$ and $c(12)$ are the total and direct correlation functions.

The common thermodynamic functions are given by
\begin{eqnarray}
E&=&\frac{1}{2}N\rho\int g(12)u(12)\rd {\bf r}_2\rd \Omega_1\Omega_2=\frac{1}{2}N\rho\int\langle g(12)u(12) \rangle\rd {\bf r}_2\,,
\\
\frac{pV}{NkT}&=&1-\frac{1}{6}\beta\rho\int \langle g(12)u'(12)R_{12} \rangle \rd {\bf r}_2\,,
\end{eqnarray}
and
{\begin{equation}
kT\frac{\partial \rho}{\partial p}=1+\rho\int h_0(12)\rd {\bf r}_2\,.
\end{equation}
In the above $E$ is the energy in excess of the kinetic energy terms, $p$ is the pressure, and $h_0(12)=g_0(12)-1$. The functions, $h_0(12)$ and $g_0(12)$, are the total and pair correlation functions, respectively.

This article gives only an outline of the field of molecular fluids.  The discussion will be restricted to molecular fluids with a hard core.  For convenience, molecular fluids can be divided further into two broad types, (1) fluids in which the hard core is spherical (the asymmetry comes from the attractive potential) and (2) fluids in which even the hard core is nonspherical.  The first class is conceptually simpler and is considered here.  Dipolar hard spheres will be considered as an example of this class.  Liquid crystals are an example of the second class and may, perhaps, be considered in a future installment.

\section{Dipolar hard spheres}

As an example of a molecular fluid, we consider the dipolar hard sphere fluid where the intermolecular potential is given by
\begin{equation}
u(12)=\left \{
\begin{array}{ll}
\infty, & R_{12}<\sigma,\\
-\frac{\mu^2}{R_{12}^3}D(12), & R_{12}>\sigma,
\end{array}
\right.
 \end{equation}
where $\mu$ and $\sigma$ are the magnitude of the dipole moment and diameter of the dipolar hard spheres,
\begin{equation}
D(12)=3({\bf \hat e}_1\cdot{\bf \hat R}_{12})({\bf \hat e}_2\cdot{\bf \hat R}_{12})-({\bf \hat e}_1\cdot{\bf \hat e}_2),
\end{equation}
where ${\bf \hat R}_{12}={\bf R}_{12}/|{\bf R}_{12}|$ is a unit vector in the direction of ${\bf R}_{12}$, ${\bf \hat e}_1$ is a unit vector in the direction
of dipole 1, and
\begin{equation}
\Delta(12)={\bf \hat e}_1\cdot{\bf \hat e}_2\,.
\end{equation}
The function $\Delta(12)$ does not appear in the intermolecular potential, except as part of $D(12)$.  However, $D(12)$ and $\Delta(12)$ contribute independently to the correlation functions.
The dipoles are assumed to be nonpolarizable.

Barker \cite{3} has proved the very useful theorem that is given in the following two equations,
\begin{equation*}
\int ({\bf \hat e}_i\cdot {\bf a}) \rd \Omega_i=0
\end{equation*}
and
\begin{equation*}
\int ({\bf \hat e}_i\cdot {\bf a})({\bf \hat e}_i\cdot {\bf b})\rd \Omega_i=\frac{1}{3}({\bf a}\cdot {\bf b}).
\end{equation*}
Note that $\bf b$ could be ${\bf \hat e}_j$, $i\neq j$.  We shall call these results Barker's theorem.

Using Barker's theorem, it is easy to show that
1, $D(12)$, and $\Delta(12)$ are orthogonal,
\begin{eqnarray}
\int D(12)\rd \Omega_1\rd \Omega_2&=&0,
\\
\int \Delta(12)\rd \Omega_1\rd \Omega_2&=&0,
\end{eqnarray}
and
\begin{equation}
\int D(12)\Delta(12)\rd \Omega_1\rd \Omega)_2=0.
\end{equation}
The normalization of 1, $D(12)$, and $\Delta(12)$ can also be obtained from Barker's theorem and is
\begin{eqnarray}
\int \rd \Omega_1\rd \Omega_2&=&1,
\\
\int D^2(12)\rd \Omega_1\rd \Omega_2&=&\frac{2}{3}\,,
\end{eqnarray}
and
\begin{equation}
\int \Delta^2(12)\rd \Omega_1\rd \Omega_2=\frac{1}{3}\,.
\end{equation}

This means that 1, $D(12)$, and $\Delta(12)$ are part of an orthogonal basis set.  Indeed, they are a subset of the spherical harmonics.
A basis set is a linearly independent set of functions with the property that any function can be expressed as a linear combination of the
members of the basis set.  One basis set for three-dimensional Euclidean space is the set of vectors in the directions of the $x$, $y$, and $z$ axes.
The space of all functions for which the spherical harmonics are the basis set has an infinite dimension.  As will be seen, in some special cases the
functions 1, $D(12)$, and $\Delta(12)$ form a complete basis set of finite (three) dimension but this is not usually the case.  A basis set need not consist of orthogonal vectors or functions.  However, it is convenient if they are orthogonal. A nonorthogonal basis set can be transformed into an orthogonal
basis set by what is called the Schmidt orthogonalization procedure.  Hence, we can expand
\begin{equation}
g(12)=g_0(R_{12})+h_D(R_{12})D(12)+h_{\Delta}(R_{12})\Delta(12)+\cdots  ,
\end{equation}
where $g_0(R_{12})$ is given by equation~(5),
\begin{equation}
h_D(R_{12})=\frac{3}{2}\int D(12)g(12)\rd \Omega_1\rd \Omega_2\,,
\end{equation}
and
\begin{equation}
h_{\Delta}(R_{12})=3\int \Delta(12)g(12)\rd \Omega_1\rd \Omega_2\,.
\end{equation}
The coefficients $g_0(R_{12}), h_{D}(12)$, and $h_{\Delta}(12)$ can be called the ``projections'' of $g(12)$ onto the basis vectors, 1, $D(12)$, and $\Delta(12)$.  The notation $h$, rather than $g$, is used for $h_D$, and $h_{\Delta}$ because they are zero when $R_{12}\rightarrow\infty$.  Note that $h_0(R)=-1$, for $R<\sigma$ but
$h_D(R)$ and $h_{\Delta}(R)$ equal 0, for $R<\sigma$.

The common thermodynamic functions are given by
\begin{eqnarray}
\frac{pV}{NkT}&=&1+y_0(\sigma)-\frac{1}{3}\beta\rho\mu^2\int\frac{h_D(R)}{R^3}\rd {\bf R},
\\
E&=&\frac{3}{2}NkT-\frac{1}{3}N\rho\mu^2\int\frac{h_D(R)}{R^3}\rd {\bf R},
\end{eqnarray}
and
\begin{equation}
kT\frac{\partial \rho}{\partial p}=1+\rho\int h_0(R)\rd {\bf R},
\end{equation}
where $y(12)$ is the background, or cavity, function, $y(12)=\exp[\beta u(12)]g(12)$ and $u(12)$ is the pair interaction.
Note that $y(12)$ is a continuous function even if $u(12)$ is discontinuous.  The functions $y_0(R)$ and $h_0(R)$ are the
spherically averaged projections of $y(12)$ and $h(12)$, respectively.

As we shall see shortly, the dielectric constant is also given by an integral involving $h_{\Delta}$.  This means that the dielectric constant and common thermodynamic functions can be obtained from $g_0$, $h_D$, and $h_{\Delta}$ even if the other projections are not known.  Of course, in general, to obtain these three projections, the other projections must be calculated.  In any case, these three projections can be called the {\it active} projections for the dipolar hard sphere fluid since they determine the common thermodynamic functions and the dielectric constant of this fluid.

\section{Simple treatments of the dielectric constant}

The simple treatments considered here are based on the concept of the {\it local} field, $E_{\mathrm{loc}}$, felt by a dipole.  This is not equal to the applied field, $E$, because of the other dipoles. Let us carve out a sphere of volume $a$, centered at a dipole.  Since the dipole-dipole interaction is long ranged, we may assume that the dielectric or polar fluid is a continuum outside this sphere.  Choose the volume of this sphere to be equal to the volume per dipole,
\begin{equation}
\frac{4\pi}{3}a^3=\frac{V}{N}=\frac{1}{\rho}\,.
\end{equation}

The average value of $\mu$ is related to $E_{\mathrm{loc}}\,$,
\begin{eqnarray}
\langle \mu\rangle &=&\langle \mu \cos \theta\rangle ,
\\
\langle \mu\rangle &=&\mu\frac{\int_0^{\pi}\cos\theta\exp(\beta\mu E_{\mathrm{loc}}\cos\theta)\sin\theta \rd \theta}{\int_0^{\pi}\exp(\beta\mu E_{\mathrm{loc}}\cos\theta)\sin\theta \rd \theta}\,.
\end{eqnarray}
As $E_{\mathrm{loc}}$ is relatively small, the exponentials in equation~(27) may be linearized.  Thus,
\begin{equation}
\langle \mu\rangle =\mu\frac{\int_0^{\pi}(1+\beta\mu E_{\mathrm{loc}}\cos\theta)\cos\theta\sin\theta \rd \theta}{\int_0^{\pi}(1+\beta\mu E_{\mathrm{loc}}\cos\theta)\sin\theta \rd \theta}\,.
\end{equation}
The integrals of the first term in the numerator and the second term in the denominator vanish.  The result is
\begin{equation}
\langle \mu\rangle =\frac{1}{3}\beta\mu^2 E_{\mathrm{loc}}\,.
\end{equation}
Our task is to calculate $E_{\mathrm{loc}}$.  We will consider two simple approaches first.

\section*{Clausius-Mossotti result for $\epsilon$}

The field inside the dielectric fluid is different from the applied field due to the polarization of this fluid.
\begin{equation}
{\bf D}=\epsilon {\bf E}={\bf E}+4\pi {\bf P},
\end{equation}
where $\bf{D}$ and $\bf{P} $ are the electric displacement and polarization vectors.  For an isotropic system, the vectors have the same direction.  Thus,
\begin{equation}
P=\frac{\epsilon -1}{4\pi}E=\rho\langle \mu\rangle .
\end{equation}

\begin{figure}[!ht]
\begin{center}
\includegraphics[width=6.5cm]{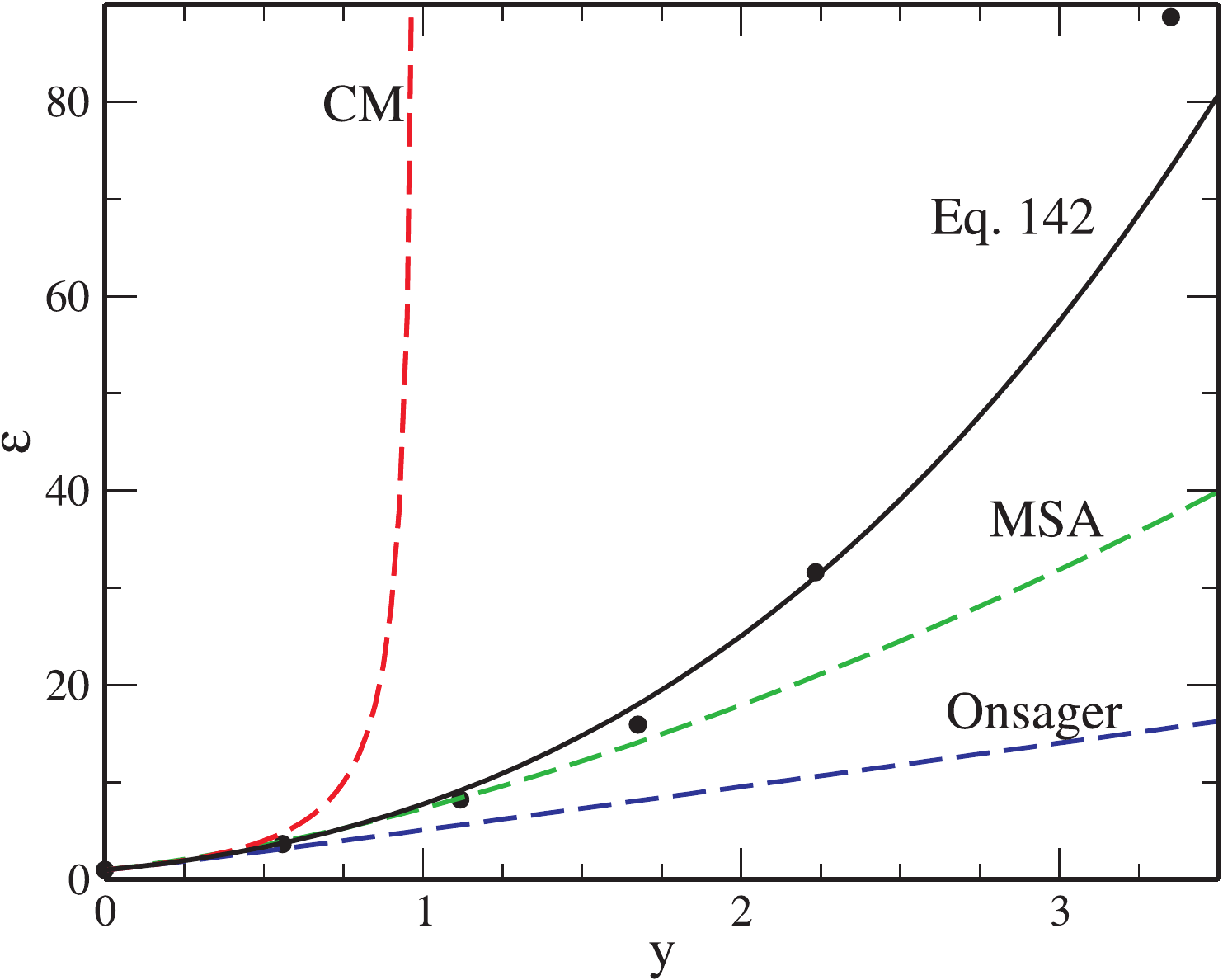}
\end{center}
\caption{Dielectric constant for the dipolar hard sphere fluid for $\rho\sigma^3=0.8$ as a function of $y$.  The solid curve
gives the result of Tani  et al. \cite{23}, equation~(142), and the broken curves give the CM \cite{6,7}, Onsager \cite{9},
and MSA \cite{12} results.  The points are simulation results \cite{8}.
}
\end{figure}
Lorentz \cite{4,5} argued that there are four contributions to $E_{\mathrm{loc}}$: (1) the applied field, $\bf{E}$; (2) the volume charge contribution of $\bf{P}$, which is zero because $\bf{P}$ is a constant and $\nabla\cdot{\bf P}=0$; (3) the surface charge contribution of $\bf{P}$ on the surface of the sphere of radius $a$; and (4) the field due to the dipole, which is independent of $\bf{E}$ and so does not contribute to $\langle \mu\rangle $.  For a surface element of this sphere at a polar angle, $\theta$, measured from the direction of $\bf P$ and $\bf E$, the area of this element is $\rd S=2\pi a^2\sin\theta \rd \theta$.  The surface charge density in the direction of the normal to the surface of the sphere at the polar angle $\theta$ due to the polarization is $P_n=P\cos\theta$.  Thus, the element of the field due to the surface polarization is $\rd E'=P_ndS/a^2$ and $E'$ is
\begin{equation}
E'=\int_0^{\pi}\cos\theta (P\cos\theta ) 2\pi\sin\theta \rd \theta= \frac{4\pi}{3}P
\end{equation}
so that
\begin{equation}
E_{\mathrm{loc}}=E+\frac{4\pi}{3}P
\end{equation}
and
\begin{equation}
\langle \mu\rangle =\frac{1}{3}\beta\mu^2\left(E+\frac{4\pi}{3}P \right),
\end{equation}
yielding
\begin{equation}
P=\frac{\frac{1}{3}\rho\beta\mu^2E}{1-\frac{4\pi}{3}\frac{1}{3}\rho\beta\mu^2}\,.
\end{equation}
Recalling that
\begin{equation}
\epsilon=1+\frac{4\pi P}{E}\,,
\end{equation}
this gives
\begin{equation}
\frac{\epsilon -1}{\epsilon+2}=\frac{4\pi}{9}\rho\beta\mu^2=y
\end{equation}
or
\begin{equation}
\epsilon=\frac{1+2y}{1-y}\,,
\end{equation}
which is the {\it Clausius-Mossotti} result \cite{6,7}.  This is not a very good result because the CM $\epsilon$ diverges when $y=1$, for which there is no experimental support.   Sometimes this problem is called the {\it polarization catastrophe}.  The CM result for $\epsilon$ is plotted and compared
with some simulation results \cite{8} in figure~1.  There is no singularity in the simulation results.

The constant $y$ is not to be confused with the background function, $y(12)$.

\section*{Onsager result for $\epsilon$}

To obtain $E_{\mathrm{loc}}$, Onsager \cite{9} solved the boundary value problem for a sphere of radius $a$ and dielectric constant equal to unity within an infinite  dielectric medium whose dielectric constant is $\epsilon$ and with an applied field $E$. Denote the potential inside and outside the sphere by $\phi_1$ and $\phi_2$, respectively.  Thus,
\begin{equation}
\nabla^2\phi_1=\nabla^2\phi_2=0.
\end{equation}
The potential and displacement are continuous across the surface of the sphere so that $\phi_1(a)=\phi_2(a)$ and $\partial \phi_1(a)/\partial R=\epsilon\partial \phi_2(a)/\partial R$.  The potential $\phi_1$ is finite inside the sphere (in particular at $R=0$) and, far from the sphere, $\phi_2=-ER\cos\theta$.  The solution of this boundary value problem is
\begin{equation}
\phi_1(R)=-\frac{3\epsilon}{2\epsilon+1}ER\cos\theta.
\end{equation}
Hence,
\begin{equation}
E_{\mathrm{loc}}=-\frac{\partial \phi_1}{\partial R}=\frac{3\epsilon}{2\epsilon+1}E.
\end{equation}
From this
\begin{equation}
\langle \mu\rangle =\frac{1}{3}\beta\mu^2\frac{3\epsilon}{2\epsilon+1}E.
\end{equation}
Using,
\begin{equation}
P=\rho\langle \mu \rangle =\frac{\epsilon-1}{4\pi}E,
\end{equation}
the dielectric constant is given by
\begin{equation}
\epsilon-1=3y\frac{3\epsilon}{2\epsilon+1}
\end{equation}
or
\begin{equation}
 \frac{(\epsilon-1)(2\epsilon+1)}{9\epsilon}=y.
\end{equation}
Explicitly, $\epsilon=[1+9y+3\sqrt{1+2y+9y^2}]/4$.  This is Onsager's formula.  It is plotted in figure~1.  This result does not diverge and is much better than the Clausius-Mossotti result.  Until Wertheim's result, this was the standard formula. Wertheim's results will now be considered.  However, some preliminary formulae are needed.

\section{Fourier transform of $h_{\Delta}(R)$}

As has been mentioned, $h(12)$ can be expanded in spherical harmonics,
\begin{equation}
 h(12)=h_0(R)+h_{\Delta}(R)\Delta(12)+h_D(R)D(12)+\cdots  .
\end{equation}
A similar expansion can be made for $c(12)$.  We will want to substitute these expressions into the OZ equation, equation~(6).  To do this it is convenient to use the Fourier transform.  The Fourier transforms of $h_0$, $c_0$, $h_{\Delta}$, and $c_{\Delta}$ are straightforward.  However, the Fourier transforms of $h_D(12)$ and $c_D(12)$ are more complicated because $D(12)$ contains $R$ and we must transform the combinations $h_D(R)D(12)$ and $c_D(R)D(12)$ as wholes.

First, recall that the Fourier transform pair is
\begin{equation}
\tilde{f}(k)=\int f(R)\exp(\ri {\bf k}\cdot {\bf r})\rd{\bf r}
\end{equation}
and
\begin{equation}
f(R)=\frac{1}{(2\pi)^3}\int \tilde{f}(k)\exp(-\ri {\bf k}\cdot {\bf r})\rd{\bf k},
\end{equation}
where $R=|{\bf r}|$ and $k=|\bf k|$.  Choose the coordinate system so that ${\bf r}=R(\sin\theta\sin\phi,\sin\theta\cos\phi,\cos\theta)$ and ${\bf k}=(0,0,k)$.  For an easier notation, define $T(12)=h_D(R)D(12)$.  After some algebra,
\begin{equation}
\tilde{T}={\bar h}_D(k)D_k(12),
\end{equation}
where
\begin{equation}
D_k(12)=3({\bf \hat e_1}\cdot {\bf \hat k})({\bf \hat e}_2\cdot {\bf\hat k})-({\bf \hat e}_1\cdot{\bf \hat e}_2)
\end{equation}
and
\begin{equation}
\bar h_D(k)=-4\pi\int_0^{\infty}R^2j_2(kR)h_D(R)\rd R,
\end{equation}
where
\begin{equation}
j_2(x)=\frac{3\sin x}{x^3}-\frac{3\cos x}{x^2}-\frac{\sin x}{x}\,.
\end{equation}
The function $\bar f(k)$ is sometimes called a {\it Hankel transform}.  Note that $\bar h_D(k)$ is not $\tilde h_D(k)$, the Fourier transform of $h_D(R)$, which is given by
\begin{equation}
\tilde h_D(k)=\frac{4\pi}{k}\int_0^{\infty}R\sin kR h_D(R)\rd R=4\pi\int_0^{\infty}R^2j_0(kR)h_D(R)\rd R.
\end{equation}
The functions $j_0(x)$ and $j_2(x)$ are spherical Bessel functions.
Equation (49) is a perfectly good result for the Fourier transform of $h_D(R)D(12)$ but it is a nuisance to have two kinds of transforms.  Thus, it is useful to define an auxiliary function,
\begin{equation}
F(R)=f(R)-3\int_R^{\infty}\frac{f(R')}{R'}\rd R',
\end{equation}
because, as is seen by straightforward integration, the Fourier transform of $F(R)$ is the Hankel transform of $f(R)$.

Thus, in summary, the Fourier transform of $T(12)=h_D(R)D(12)$ is given by
\begin{equation}
\tilde{T}(12)=\tilde{H}_D(k)D_k(12),
\end{equation}
where $D_k(12)$ is given by equation~(50) and
\begin{equation}
H_D(R)=h_D(R)-3\int_R^{\infty}\frac{h_D(R')}{R'}\rd R'.
\end{equation}
The inverse of the last equation is
\begin{equation}
h_D(R)=H_D(R)-\frac {3}{R^3}\int_0^RH_D(R')R'^2\rd R'.
\end{equation}

This auxiliary function has another interesting property.  If $f(R)$ is long ranged, $F(R)$ is short ranged.  For example,
if
\begin{equation}
f(R)=\left \{
\begin{array}{ll}
0, & R<\sigma, \\
-\frac{1}{R^3}\,, & R>\sigma,
\end{array}
\right.
 \end{equation}
then
\begin{equation}
F(R)=\left \{
\begin{array}{ll}
-\frac{1}{\sigma^3}\,, & R<\sigma, \\
0, & R>\sigma.
\end{array}
\right.
\end{equation}
This property can be exploited to evaluate integrals of long ranged functions which would be difficult if direct integration were attempted.

\section{Fourier transform of the OZ equation for dipolar hard spheres}

It has been seen that
\begin{equation}
h(12)=h_0(R)+h_{\Delta}(R)\Delta(12)+h_D(R)D(12)+\cdots
\end{equation}
and
\begin{equation}
c(12)=c_0(R)+c_{\Delta}(R)\Delta(12)+c_D(R)D(12)+\cdots .
\end{equation}
Hence,
\begin{equation}
\tilde{h}(12)=\tilde{h}_0(k)+\tilde{h}_{\Delta}(k)\Delta(12)+\tilde{H}_D(k)D_k(12)+\cdots
\end{equation}
and
\begin{equation}
\tilde{c}(12)=\tilde{c}_0(k)+\tilde{c}_{\Delta}(k)\Delta(12)+\tilde{C}_D(k)D_k(12)+\cdots .
\end{equation}

\begin{table}[ht]
\caption{Wertheim's ``multiplication'' table.}
\begin{center}
\begin{tabular}{c|ccc}
\hline\hline
&1&$\Delta(23)$&$D_{K}(23)$\\
\hline
1&1&0&0\\[1ex]
$\Delta(13)$&0&$\frac13 \Delta(12)$&$\frac13 D_{K}(12)$\\[1ex]
$D_{K}(13)$&0&$\frac13 D_{K}(12)$&$\frac13 \left[D_{K}(12)+2\Delta(12)\right]$\\[1ex]
\hline\hline
\end{tabular}
\label{table1}
\end{center}
\end{table}
To take the transform of the convolution in the OZ relation, we must evaluate integrals of the form
\begin{equation*}
\int \Delta(13)D(23)\rd \Omega_3\,.
\end{equation*}
To do this, Wertheim's ``multiplication table'', which is given in table~\ref{table1}, is required.  This multiplication table
is easily obtained using Barker's theorem.   Using this table, the transform of the OZ equation may be obtained.
Since 1, $\Delta(12)$, and $D(12)$ are orthogonal, we can equate coefficients.  Thus,
\begin{equation}
\tilde{h}_0=\tilde{c}_0+\rho\tilde{h}_0\tilde{c}_0+\cdots ,
\end{equation}
\begin{equation}
\tilde{h}_{\Delta}=\tilde{c}_{\Delta}+\frac{1}{3}\rho(\tilde{h}_{\Delta}\tilde{c}_{\Delta}+2\tilde{H}_D\tilde{C}_D)+\cdots ,
\end{equation}
and
\begin{equation}
\tilde{H}_D=\tilde{C}_D+\frac{1}{3}\rho(\tilde{h}_{\Delta}\tilde{C}_D+\tilde{H}_D\tilde{c}_{\Delta}+\tilde{H}_D\tilde{C}_D)+\cdots ,
\end{equation}
with similar equations for the transforms of the higher order terms.

We know that
\begin{eqnarray}
&&h_0(R)=-1, \qquad R<\sigma,\\
&&h_0(R)\rightarrow 0,\qquad R\rightarrow\infty,\\
&&c_0(R)\rightarrow 0,\qquad R\rightarrow\infty,
\end{eqnarray}
\begin{eqnarray}
&&h_{\Delta}(R)=0, \qquad R<\sigma,\\
&&h_{\Delta}(R)\rightarrow 0,\qquad R\rightarrow\infty,\\
&&c_{\Delta}(R)\rightarrow 0,\qquad R\rightarrow\infty,
\end{eqnarray}
and
\begin{eqnarray}
&&h_D(R)=0, \qquad R<\sigma,\\
&&h_D(R)\rightarrow 0,\qquad R\rightarrow\infty,\\
&&c_D(R)\rightarrow \frac{\beta\mu^2}{R^3}\,,\qquad R\rightarrow\infty.
\end{eqnarray}

Equations (73)--(75) are fine but we are interested in $H_D(R)$ and $C_D(R)$ rather than in $h_D(R)$ and $c_D(R)$.  It is
easy to show that
\begin{eqnarray}
&&H_D(R)=-3K, \qquad R<\sigma,\\
&&H_D(R)\rightarrow 0,\qquad R\rightarrow\infty,\\
&&C_D(R)\rightarrow 0, \qquad R\rightarrow\infty,
\end{eqnarray}
where
\begin{equation}
K=\int_{\sigma}^{\infty}\frac{h_D(R)\rd R}{R}\,.
\end{equation}
The parameter, $K$, is independent of $R$ but depends on $T$, $\rho$, $\epsilon$, etc., and is not known until the
problem is solved.

We can establish an interesting result for $\tilde{C}_D(0)$.  We know that
\begin{equation}
c_D(R)=C_D(R)-\frac{3}{R^3}\int_0^RC_D(R')R'^2\rd R'.
\end{equation}
Using equations~(75) and (78) it follows that
\begin{equation}
\frac{\beta\mu^2}{R^3}=-\frac{3}{R^3}\int_0^{\infty}C_D(R')R'^2\rd R'.
\end{equation}
From this, we have
\begin{equation}
-\frac{1}{3}\rho \tilde{C}_D(0)=y.
\end{equation}

\section{Some exact results for $\epsilon$}

Onsager's expression is a special case of the exact result \cite{10,11}
\begin{equation}
\frac{(\epsilon-1)(2\epsilon+1)}{9\epsilon}=\frac{4\pi\beta\rho}{9}\frac{\langle M^2 \rangle}{N},
\end{equation}
where $\bf M$ is the total dipole moment of the dielectric.  Write this as
\begin{equation}
\frac{(\epsilon-1)(2\epsilon+1)}{9\epsilon}=yg_K\,.
\end{equation}
The parameter $g_K$ is called the Kirkwood $g_K$ factor.  The $g_K$ factor can be written as an integral,
\begin{equation}
g_K=\frac{\langle M^2 \rangle}{N\mu^2}
=1+N \langle {\bf \hat{e}}_1\cdot{\bf\hat{e}}_2 \rangle
\end{equation}
yielding
\begin{equation}
g_K=1+\frac{1}{3}\rho\int h_{\Delta}(R)\rd {\bf R}=1+\frac{1}{3}\rho\tilde{h}_{\Delta}(0)
\end{equation}
so that
\begin{equation}
\frac{(\epsilon-1)(2\epsilon+1)}{9\epsilon}=y\left[1+\frac{1}{3}\rho \tilde{h}_{\Delta}(0)\right].
\end{equation}
The Onsager approximation consists in neglecting the contribution of $h_{\Delta}(R)$.

Some other interesting exact results for $\epsilon$ can be obtained using the OZ
equation given above.  We can use the truncated versions of the expressions for $h_0$, $h_{\Delta}$, and
$h_D$, namely,
\begin{equation}
\tilde{h}_0=\tilde{c}_0+\rho\tilde{h}_0\tilde{c}_0\,,
\end{equation}
\begin{equation}
\tilde{h}_{\Delta}=\tilde{c}_{\Delta}+\frac{1}{3}\rho(\tilde{h}_{\Delta}\tilde{c}_{\Delta}+2\tilde{H}_D\tilde{C}_D),
\end{equation}
and
\begin{equation}
\tilde{H}_D=\tilde{C}_D+\frac{1}{3}\rho(\tilde{h}_{\Delta}\tilde{C}_D+\tilde{H}_D\tilde{c}_{\Delta}+\tilde{H}_D\tilde{C}_D).
\end{equation}
The missing terms do not contribute.  Solving the truncated equation~(89) for
$\tilde{h}_{\Delta}$ gives
\begin{equation}
 \tilde{h}_{\Delta}=\frac{\tilde{c}_{\Delta}+\frac{2}{3}\rho\tilde{H}_D\tilde{C}_D}{1-\frac{1}{3}\rho \tilde{c}_{\Delta}}\,.
\end{equation}
Solving equation~(90) for $\tilde{H}_D$ gives, for $k=0$,
\begin{equation}
\tilde{H}_D(0)=\frac{\tilde{C}_D(0)}{(x-y)(x+2y)}\,,
\end{equation}
where $x=1-\frac{1}{3}\rho \tilde{c}_{\Delta}(0)$ and equations~(82) and (91) have been used.

Equation (91) can be rewritten as
\begin{equation}
1+\frac{1}{3}\rho\tilde{h}_{\Delta}=\frac{\frac{2}{9}\rho^2\tilde{H}_D\tilde{C}_D}{1-\frac{1}{3}\rho\tilde{c}_{\Delta}}\,.
\end{equation}
Thus,
\begin{equation}
x\left[1+\frac{1}{3}\rho h_{\Delta}(0)\right]=1-\frac{2}{3}\rho \tilde{H}(0)y.
\end{equation}
Using equation~(92) yields
\begin{equation}
\frac{x+y}{(x-y)(x+2y)}=1+\frac{1}{3}\rho\tilde{h}_{\Delta}(0)=\frac{(\epsilon-1)(2\epsilon+1)}{9y\epsilon}\,.
\end{equation}
The solution of this equation can be verified to be
\begin{equation}
x=y\frac{\epsilon+2}{\epsilon-1}=1-\frac{1}{3}\rho\tilde{c}_{\Delta}(0)
\end{equation}
or
\begin{equation}
\frac{\epsilon-1}{\epsilon+2}=\frac{y}{1-\frac{1}{3}\rho \tilde{c}_{\Delta}(0)}\,.
\end{equation}
Hence, the Clausius-Mossotti result is obtained by neglecting $c_{\Delta}(R)$.

Finally
\begin{equation}
\frac{1}{3}\rho\tilde{H}_D(0)=-\frac{y}{(x-y)(x+2y)}
\end{equation}
or
\begin{equation}
\frac{1}{3}\rho\tilde{H}_D(0)=-\frac{(\epsilon-1)^2}{9\epsilon y}.
\end{equation}
These three routes to $\epsilon$ may not be consistent for a given approximation.
However, they will be consistent if the OZ relation is satisfied.

Note also, that we have obtained exact expressions for $\tilde{h}_{\Delta}(0)$, $\tilde{H}_D(0)$,
$\tilde{c}_{\Delta}(0)$, and $\tilde{C}_D(0)$!

\section{The mean spherical approximation for the dipolar hard sphere fluid}

Because the MSA is a linearized approximation, 1, $\Delta$, and $D$ are a complete
basis set for the MSA.  Thus, equations~(88)--(90) can be employed.  The MSA is
\begin{eqnarray}
h_0&=&-1,\qquad  R<\sigma,
\\
c_0&=&0, \qquad R>\sigma,
\\
h_{\Delta}&=&0, \qquad R<\sigma,
\\
c_{\Delta}&=&0, \qquad R>\sigma,
\end{eqnarray}
and
\begin{eqnarray}
H_D&=&-3K, \qquad R<\sigma,
\\
C_D&=&0, \qquad R>\sigma.
\end{eqnarray}
These equations were first obtained and solved by Wertheim \cite{12}.

The first thing to note is that $h_0$ and $c_0$ are decoupled from
equations~(89) and (90).  Equations (100) and (101) are the Percus-Yevick (PY)
approximation for a hard sphere (HS) fluid.  Thus,
\begin{equation}
h_0(R)=h_{\mathrm{HS}}^{\mathrm{PY}}(R).
\end{equation}
Algorithms for
calculating $h_{\mathrm{HS}}^{\mathrm{PY}}(R)=g_{\mathrm{HS}}^{\mathrm{PY}}(R)-1$, that are based on the formulae of Thiele \cite{13} and
Wertheim \cite{14,15}, have been given previously \cite{16,17}. The other two equations may be solved
by introducing the new functions,
\begin{equation}
h_+(R)=\frac{1}{3K}\left[H_D(R)+\frac{1}{2}h_{\Delta}(R)\right]
\end{equation}
and
\begin{equation}
h_-(R)=\frac{1}{3K}[H_D(R)-h_{\Delta}(R)].
\end{equation}

After a little algebra, the decoupled equations,
\begin{equation}
 \tilde{h}_+=\tilde{c}_+ +2K\rho\tilde{h}_+\tilde{c}_+
\end{equation}
and
\begin{equation}
 \tilde{h}_-=\tilde{c}_- -K\rho\tilde{h}_-\tilde{c}_-
\end{equation}
follow.  The MSA approximation consists of
\begin{eqnarray}
h_+&=&-1, \qquad R<\sigma,
\\
c_+&=&0, \qquad R>\sigma
\end{eqnarray}
and
\begin{eqnarray}
h_-&=&-1, \qquad  R<\sigma,
\\
c_-&=&0,  \qquad R>\sigma.
\end{eqnarray}

Hence,
\begin{equation}
h_+(R;\rho)=h_{\mathrm{HS}}^{\mathrm{PY}}(R;2K\rho)
\end{equation}
and
\begin{equation}
h_-(R;\rho)=h_{\mathrm{HS}}^{\mathrm{PY}}(R;-K\rho),
\end{equation}
where $h_{\mathrm{HS}}^{\mathrm{PY}}(R;\rho)$ are the PY hard sphere correlation functions.  The equations for $c_+$ and $c_-$ are
similar.  The contact values of $h_+(R;\rho)$ and $h_(R;\rho)$ are given by
\begin{equation}
h_+(\sigma;\rho)=K\eta\frac{5-4K\eta}{(1-2K\eta)^2}
\end{equation}
and
\begin{equation}
h_-(\sigma;\rho)=-K\eta\frac{5+2K\eta}{2(1+K\eta)^2}\,,
\end{equation}
where $\eta=\pi\rho\sigma^3/6$.

The algorithms of Smith  et al. \cite{16,17} can be used.  These algorithms are robust and,
with a small change, give sensible results, even for the negative densities required by equation~(116).  The required
change is that the one line in the program where a cube root of a quantity involving $\eta$ is taken, the
instruction should be changed so that when $\eta$ is negative, the absolute value of $\eta$ is used and the
resulting cube root is multiplied by $-1$.

Thus,
\begin{equation}
h_{\Delta}(R)=2K\left[h_{\mathrm{HS}}^{\mathrm{PY}}(R;2K\rho)-h_{\mathrm{HS}}^{\mathrm{PY}}(R;-K\rho)\right]
\end{equation}
and
\begin{equation}
H_{D}(R)=2K\left[h_{\mathrm{HS}}^{\mathrm{PY}}(R;2K\rho)+\frac{1}{2}h_{\mathrm{HS}}^{\mathrm{PY}}(R;-K\rho)\right].
\end{equation}
The contact values of $h_{\Delta}(R)$ and $H_D(R)$ follow from equations~(117) and (118) together with
\begin{equation}
g_{\mathrm{HS}}^{\mathrm{PY}}(\sigma;\rho)=\frac{1+\eta/2}{(1-\eta)^2}\,.
\end{equation}
The function that we want is $h_D(R)$, not $H_D(R)$.  This can be calculated from equation~(57).  In particular, the contact value of $h_D(R)$ is
\begin{equation}
h_D(\sigma)=H_D(\sigma)+3K.
\end{equation}

The parameter $K$ is not yet specified but we are in a position to do so now.
Using equation~(82) and
\begin{equation}
1-\rho\tilde{c}_{\mathrm{HS}}^{\mathrm{PY}}(0)=\frac{(1+2\eta)^2}{(1-\eta)^4}
\end{equation}
yields
\begin{equation}
\frac{(1+4K\eta)^2}{(1-2K\eta)^4}-\frac{(1-2K\eta)^2}{(1+K\eta)^4}=3y,
\end{equation}
which specifies $K$, which has been renormalized so that it is dimensionless.
Note that $0<K\eta<1/2$.  When $K\eta=0$, $y=0$ and when $K\eta=1/2$, $y=\infty$.

\begin{figure}[h]
\begin{center}
\includegraphics[width=10cm]{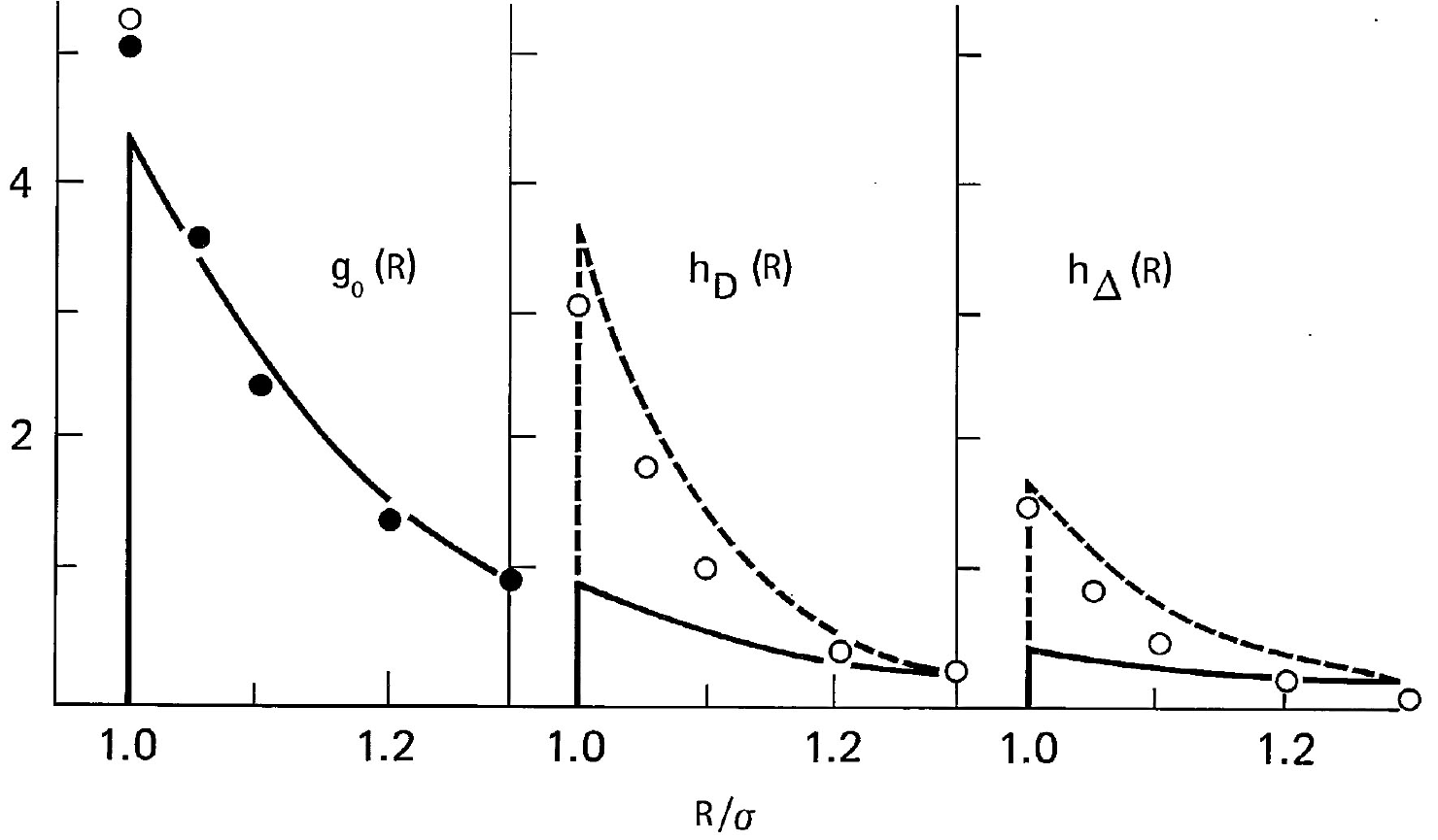}
\end{center}
\caption{Correlation functions for the dipolar hard sphere fluid
for $\rho\sigma^3=0.9$. The points given by solid and open circles are
the simulation results \cite{18,19} for $\beta\mu^2=0$ (hard spheres) and $\beta\mu^2=1$, respectively.
The solid and broken curves give the results of the MSA and equations~(125)
and (126), respectively, for $\beta\mu^2=1$.
}
\label{fig2}
\end{figure}
The correction functions for the dipolar hard sphere fluid that
follow from the MSA are plotted and compared with simulation results \cite{18,19} in figure~2
for a representative case.  The MSA gives fairly accurate results for $g_0(R)$.  The
simulation results for $g_0(R)$ for dipolar hard spheres are very nearly equal to
those for hard spheres but are slightly larger.  Hence, one prediction of the
MSA is that $g_0(R)$ for dipolar hard spheres is independent of the magnitude of
the dipole moment and is equal to the radial distribution function of a
hard sphere fluid.  This prediction is not exact but is quite well satisfied by the simulation
results.  However, the MSA results for $h_D(R)$ and
$h_{\Delta}(R)$ are rather poor.  Interestingly, the approximations, called LEXP,
\begin{equation}
h_D(R)=g_{\mathrm{HS}}^{\mathrm{PY}}(R)h_D^{\mathrm{MSA}}(R)
\end{equation}
and
\begin{equation}
h_{\Delta}(R)=g_{\mathrm{HS}}^{\mathrm{PY}}(R)h_{\Delta}^{\mathrm{MSA}}(R),
\end{equation}
are much better.

\section{MSA thermodynamic functions}

Using the compressibility route, the thermodynamic functions are as follows:
\begin{equation}
\beta\frac{\partial p}{\partial\rho}=\frac{(1+2\eta)^2}{(1-\eta)^4}\,.
\end{equation}
When the compressibility equation is used, this is a very poor result since the MSA incorrectly predicts, that there is no contribution
from the dipolar part of the intermolecular potential.

Using the pressure route,
\begin{equation}
\frac{pV}{NkT}=\frac{1+2\eta+3\eta^2}{(1-\eta)^2}-\frac{4\pi}{3}\beta\rho\mu^2\int_{\sigma}^{\infty}\frac{h_D(R)}{R}\rd R,
\end{equation}
which becomes
\begin{equation}
\frac{pV}{NkT}=\frac{1+2\eta+3\eta^2}{(1-\eta)^2}-3Ky,
\end{equation}
and, using the energy route,
\begin{equation}
\frac{E}{NkT}=\frac{3}{2}-3Ky.
\end{equation}

\begin{figure}[ht]
\begin{center}
\includegraphics[width=7cm]{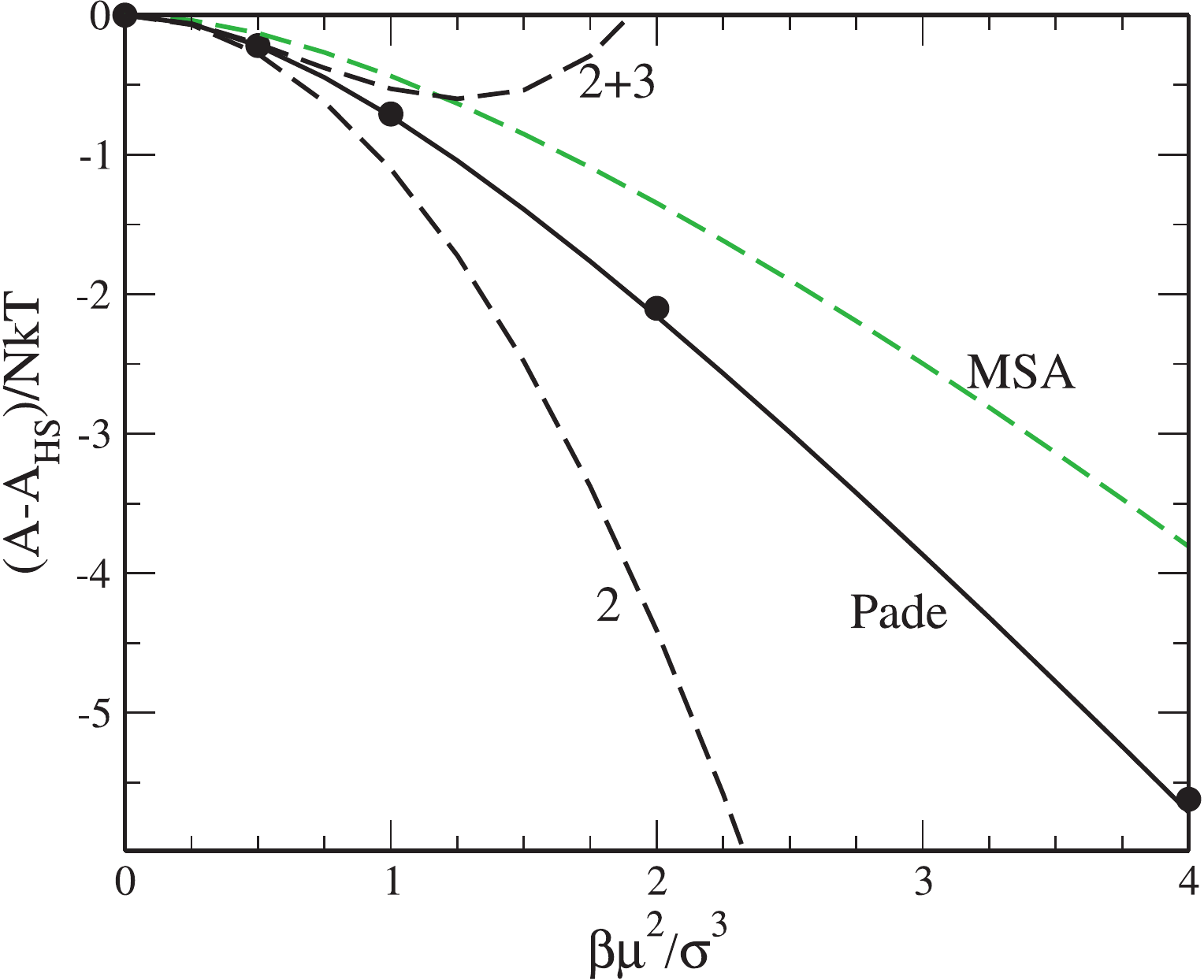}
\end{center}
\caption{Free energy of the dipolar hard sphere fluid for $\rho\sigma^3=0.9$ as a
function of $\beta\mu^2/\sigma^3$.  The points are simulation results \cite{19}.  The
dashed curves marked MSA, 2, and 2+3 give the results of the MSA and
perturbation theory when truncated after 2 and 3 terms, respectively.  The solid
curve gives the results of the Pad\'{e} extrapolation of Rushbrooke  et al. \cite{24}.
}
\label{fig3}
\end{figure}
One interesting characteristic of the MSA is that the energy can be
integrated analytically to give the (energy equation) free energy and this free
energy can be differentiated to yield the (energy equation) pressure.  After
a little algebra, the results are
\begin{equation}
\frac{A-A_{\mathrm{HS}}}{NkT}=-K^2\eta\left[8\frac{(1+K\eta)^2}{(1-2K\eta)^4}+\frac{(2-K\eta)^2}{(1+K\eta)^4}\right]
\end{equation}
and
\begin{equation}
\frac{p-p_{\mathrm{HS}}}{\rho kT}=K\eta^2\left[8\frac{(1+K\eta)^2}{(1-2K\eta)^4}+\frac{(2-K\eta)^2}{(1+K\eta)^4}\right]-3Ky,
\end{equation}
where $A_{\mathrm{HS}}$ and $p_{\mathrm{HS}}$ are the hard sphere free energy and pressure, respectively.  Note that $K$
has been renormalized so that it is dimensionless.  The free energy that
results from the MSA is plotted and compared with simulation results \cite{20} in figure~3.  As is usually
the case, the thermodynamics obtained from the energy equation are more accurate than those obtained from
the compressibility or pressure equations.

\section{MSA dielectric constant}

The dielectric constant can be calculated by the three routes given above.  All three routes yield the
same expression for $\epsilon$.  For example, starting with equation~(97), we obtain
\begin{equation}
3y\frac{\epsilon+2}{\epsilon-1}=\frac{(1+4K\eta)^2}{(1-2K\eta)^4}+2\frac{(1-2K\eta)^2}{(1+K\eta)^4}\,.
\end{equation}
Solving for $\epsilon$ gives
\begin{equation}
\epsilon=\frac{(1+4K\eta)^2(1+K\eta)^4}{(1-2K\eta)^6}\,.
\end{equation}
Some MSA results for $\epsilon$ are plotted in figure~1.  The
agreement of the MSA result with the simulation results is better than for
the CM and Onsager theories but the MSA results are still
too small.

Expanding the MSA expression for $\epsilon$ gives
\begin{equation}
\frac{\epsilon -1}{\epsilon+2}=y-\frac{15}{16}y^3+\cdots ,
\end{equation}
which is correct to order $y^3$.  By contrast, the CM result is
\begin{equation}
\frac{\epsilon -1}{\epsilon+2}=y
\end{equation}
and the Onsager result is, on expansion,
\begin{equation}
\frac{\epsilon -1}{\epsilon+2}=y-2y^3+\cdots .
\end{equation}
There is no term of order $y^3$ in the CM theory.  The Onsager coefficient of the $y^3$ term is too negative.

Of course, approximations that are better than the MSA approximation can be used.  For example, Fries and Patey~\cite{21}
used the hypernetted chain (HNC) approximation.  The HNC results are better than the MSA results but the calculations
are lengthy and, in contrast to the theories considered here, do not yield analytic results. It is to be noted
that other combinations of $\beta$, $\mu$, and $\rho$, besides $y$, appear when the HNC approximation is employed.
This is true of other more general theories, for example the perturbation theory that is considered below.

\section{Perturbation theory for dipolar hard spheres}

Perturbation theory has been found to be very successful for simple fluids.  It is natural
to wonder if perturbation theory might also be useful for a polar fluid.  The answer is
yes but some qualifications are necessary.

By expanding the free energy in powers of $\beta$, the following result is obtained
\begin{equation}
A=A_{\mathrm{HS}}+\beta^2\mu^4A_2+\beta^3\mu^6A_3 + \cdots ,
\end{equation}
where $A_{\mathrm{HS}}$ is the hard sphere free energy.  The quantities $A_2$ and $A_3$ are given by
\begin{equation}
\frac{A_2}{NkT}=-\frac{1}{4}\rho\int\frac{\langle D^2(12)\rangle}{R^6}g_{\mathrm{HS}}(R)\rd {\bf R}=-\frac{1}{6}\rho\int \frac{g_{\mathrm{HS}}(R)}{R^6}\rd {\bf R}
\end{equation}
and
\begin{equation}
\frac{A_3}{NkT}=\frac{1}{6}\rho^2\int\frac{\langle D(12)D(13)D(23)\rangle }{(R_{12}R_{13}R_{23})^3}g_{\mathrm{HS}}(123)\rd {\bf r}_2\rd {\bf r}_3=\frac{1}{54}\rho^2I_{ddd}\,,
\end{equation}
where $g_{\mathrm{HS}}(R)$ and $g_{\mathrm{HS}}(123)$ are the pair and triplet  distribution functions of the hard sphere fluid, and
\begin{equation}
I_{ddd}=\int\frac{1+3\cos\theta_1\cos\theta_2\cos\theta_3}{(R_{12}R_{13}R_{23})^3}g_{\mathrm{HS}}(123)\rd {\bf r}_2\rd {\bf r}_3\,,
\end{equation}
where the $\theta_i$ are the three interior angles of the triangle formed by the three sides, $R_{ij}$.  The angle $\theta_1$
is the angle opposite the side $R_{23}$, etc. Barker's theorem has been used to perform/simplify the angular integrations.
The term of order $\beta$ vanishes on angular integration, as do some of the terms of order $\beta^2$ and $\beta^3$ that formally contribute.
Barker  et al. \cite{22} and Tani  et al. \cite{23} have calculated $I_{ddd}$ by simulation and direct integration via the superposition
approximation, $g_{\mathrm{HS}}(123)=g_{\mathrm{HS}}(12)g_{\mathrm{HS}}(13)g_{\mathrm{HS}}(23)$.  A numerical fit of their results is
given by
\begin{equation}
I_{ddd}=\frac{5\pi^2}{3}\sigma^6\frac{1+1.12754\rho^*+0.56192\rho^{*2}}{1-0.05495\rho^*+0.13332\rho^{*2}}\,,
\end{equation}
where $\rho^*=\rho\sigma^3$.

As is seen in figure~3, this truncated series gives poor results.  However the Pad\'{e},
\begin{equation}
A=A_{\mathrm{HS}}+\beta^2\mu^4\frac{A_2}{1-\beta\mu^2\frac{A_3}{A_2}}\,,
\end{equation}
that was proposed by Rushbrooke  et al. \cite{24}, gives excellent agreement with the simulation results.
A Pad\'{e} tends to work best for alternating series.  For example, the series $1-1+1-1+\cdots $ is
summed correctly to 1/2 by a Pad\'{e}.
Patey and Valleau \cite{20} refer to the Pad\'{e} results as ``absurdly successful''.  This is
meant as a positive comment and is a fair observation.
Unfortunately, a Pad\'{e} does not work well for the correlation functions.  Some thoughts about the
development of approximations that are consistent with equations~(143) have been considered by Barker and
Henderson \cite{25}.  However, nothing much has come of these efforts.

The dielectric constant can be calculated from
\begin{equation}
\frac{(\epsilon-1)(2\epsilon+1)}{9\epsilon}=y\left[1+\frac{9I_{dd\Delta}}{16\pi^2}+\cdots \right]
\end{equation}
where \cite{22,23}
\begin{equation}
I_{dd\Delta}=\int\frac{3\cos^2\theta_3-1}{(R_{13}R_{23})^3}g_{\mathrm{HS}}(123)\rd {\bf r}_2\rd {\bf r}_3=\frac{17\pi^2}{9}\sigma^6\frac{1-0.93952\rho^*+0.36714\rho^{*2}}{1-0.92398\rho^*+0.23323\rho^{*2}}\,.
\end{equation}
\vspace{1ex}
This gives poor results, even with a Pad\'{e}.  However, the  direct expansion, due to Tani  et al.~\cite{23},
\begin{equation}
\epsilon=1+3y+3y^2+3y^3\left(\frac{9I_{dd\Delta}}{16\pi^2}-1\right),
\end{equation}
gives very good results, as is in figure~1.

Perturbation theory can be recast by subtracting the MSA contributions from the perturbation terms
and writing the perturbation theory as a series of corrections to the MSA.  This was done by Henderson
 et al.~\cite{26}.  The results are similar to those of the perturbation theory considered here.

\section{A few remarks}

The mean spherical approximation and perturbation theory are pleasing
extensions of the classic theories of Clausius and Mossotti and Onsager for polar fluids.
Not only do they make more accurate predictions for the dielectric constant of a
polar fluid but they predict the other thermodynamic properties of polar fluids.
Many of these ideas are applicable to polar fluids with a dispersion
interaction.  The simplest system of the kind is the dipolar Yukawa fluid.  Szalai
 et al.~\cite{27} and Mate  et al.~\cite{28} have considered this model to be polar fluid.

The molecules considered here are unpolarizable dipoles.  Onsager considered
polarizable dipoles.  Valisk\'{o}  et al. have generalized some of the
expressions presented here for polarizable dipoles and have made simulations
for a polarizable dipolar hard sphere fluid.

The author had hoped to
include his lecture notes on liquid crystals as an example of a molecular
fluid with a nonspherical hard core.  However, despite some searching, these
have not been found.  If they do come to light, they can form a fourth part of
this series and the ``concerto'' with three movements can become a ``symphony'' with four movements.

\section*{Acknowledgements}

The lecture notes on which this paper and~\cite{1,2} are based were written in 1988 when the author was in Mexico City as the Manuel Sandoval Vallarta visiting professor of physics at the Universidad Autonoma Metropolitana, Iztapalapa Campus.  There was a strike at the university during the month of February.  The author used this time to good purpose.  Each day was spent writing these notes in the study of the late Professor Manuel Sandoval Vallarta.  The purpose of this article is to record these notes by publication for the benefit of students and to record a few new results.  Marcelo Lozada-Cassou and Luis Mier y Teran arranged this visiting professorship.  The friendship of Marcelo, Luis, and Fernando del Rio is remembered with pleasure.  The author is grateful for their efforts and for the hospitality of Sra Sandoval Vallarta.  He is also grateful to the strikers for giving him this free time. Gren  Patey, Dezs\H{o} Boda, and Andrij Trokhymchuk have helped with the preparation of this manuscript.  However, the author is solely responsible for any errors.

\ukrainianpart

\title{Деякі прості результати для властивостей полярних плинів}

\author{Д. Гендерсон}

\address{Відділ хімії та біохімії, університет Брігема Янга, Прово, штат Юта 84602}

\makeukrtitle

\begin{abstract}
Підсумовується лекційний матеріал автора, присвячений кореляційним функціям і термодинаміці простого полярного плину. Особлива увага приділяється дипольному плину твердих сфер і середньо-сферичному наближенню, а також зв'язку цих результатів із формулами Клаузіуса-Мосотті та Онзагера для діелектричної сталої. Попередні викладки із цих лекцій, Condens. Matter Phys., 2009, \textbf{12}, 127; ibid., 2010, \textbf{13}, 13002, містили результати, які не були загальновідомими. Є надія що ця третя і, ймовірно, остання викладка буде такою ж корисною, об'єднуючи кілька результатів і роблячи їх доступними для ширшої аудиторії, а також представляючи кілька нових результатів.

\keywords кореляційні функції, полярні плини, термодинамічні функції, діелектрична стала
\end{abstract}

\end{document}